\pdfoutput=1
\documentclass[a4paper]{article}

\usepackage{multirow}
\usepackage{array}
\usepackage[a4paper]{geometry}
\usepackage{lscape}
\usepackage{longtable}
\usepackage{booktabs} % For formal tables
\usepackage{graphicx}
\usepackage{multirow}
\usepackage{multicol}
\usepackage{float}
\usepackage{xcolor}
\usepackage{xcolor,colortbl}
\usepackage[hyphens]{url}
\usepackage[english]{datetime2}

\date{}
\begin{document}
	\title{A report on personally identifiable sensor data from smartphone devices}
	\author{Marios Fanourakis \\ CUI, Quality of Life Lab \\ University of Geneva, Switzerland}
	\maketitle
	
	\begin{abstract}
		An average smartphone is equipped with an abundance of sensors to provide a variety of vital functionalities and conveniences. The data from these sensors can be collected in order to find trends or discover interesting correlations in the data but can also be used by nefarious entities for the purpose of revealing the identity of the persons who generated this data. In this paper, we seek to identify what types of sensor data can be collected on a smartphone and which of those types can pose a threat to user privacy by looking into the hardware capabilities of modern smartphone devices and how smartphone data is used in the literature. We then summarize some implications that this information could have on the GDPR.
	\end{abstract}
	
	\section{Introduction}
	Data can be obtained through many types of sensors or surveys. Already, the average mobile device includes several sensors as a standard feature: accelerometer (to know when the screen rotates), compass (for positioning), GPS (for positioning), light (to adjust the display brightness), audio (microphone), image (camera), and others. Other devices may include different types of sensors like temperature, air quality, heart rate, etc. Moreover, a mobile phone with such sensors roams with its owner, and can be used to collect context information on their behalf.
	
	An average smartphone is equipped with an abundance of sensors to provide a variety of vital functionalities and conveniences. For example, the basic telephony antenna which enables the smartphone to connect to the cellular network, or the ambient light sensor which helps to automatically adjust the screen brightness to a comfortable level. The data that these sensors provide pose no threat when used for their intended purpose. 
	
	With the advent of crowd sensing, this data is collected indiscriminately in order to find trends or discover interesting correlations in the data and are often kept in large databases where malicious entities can use it for nefarious purposes by revealing the identity of the persons who generated this data. For this reason, there has been a noticeable effort in the research community to develop methods and strategies to protect the privacy of the users while still being able to collect usable data from them. These methods can introduce limitations in the utility of the data and in, some cases, a non-negligible overhead in the overall data collection and data mining processes, therefore, it is advantageous to know which data has the potential to be a threat to a user's privacy so that only that data and no other is treated with the privacy-preserving methods that have been developed.
	
	In this paper, we seek to identify what types of sensor data can be collected on a smartphone and which of those types can pose a threat to user privacy. We identify the data types by first looking at the hardware specifications of a typical smartphone and then looking at the Android API to see what information can be retrieved from this hardware. To determine the threat level of each type we look into the literature for how this data can be used (for example, in behavioural biometrics, inference attacks, behaviour modeling, etc.). Considering the large scope of the topic we choose to focus on novel or recent work which is based on or improves upon older work. % and if the literature is lacking or inconclusive we perform our own investigation by collecting samples and analyzing them.
	
	\section{Smartphone Data Types and Privacy}
	\label{sec:dataThreats}
	After analyzing hardware information of popular smartphones by Samsung, \linebreak Google, LG, HTC, and Huawei we present in this section the most common sensors and data available. The information is presented in no particular order.
	
	For certain data types we equate their potential for a privacy breach with their usefulness as authentication modalities. Our reasoning is that if a particular data type is unique enough to be used for authentication purposes, then it is certainly unique enough to identify a person in a large dataset. Furthermore, if a particular data type is significantly correlated with another data type which was evaluated to be a privacy threat, then we conclude that this particular data type must also be a privacy threat to some level.
	
	\paragraph{Sampling Types}
	For each data type we also note the type of sampling that is required to derive useful information. In table \ref{table:samplingTypes} we define several sampling types that can be used to collect information from sensors. We can use these sampling types to qualify the threat-sensitivity of a particular sensor's data. For example, if some sensor only requires Type A sampling, where a single sample is enough to derive some feature, then we can conclude that it a sensor which can easily compromise an individual's privacy and should avoid sharing any of its data or be extremely selective over which data to share and to whom. On the other hand, if a sensor requires Type F sampling, several samples can be shared without compromising the privacy of an individual. 
	\begin{table}[H]
		\resizebox{\textwidth}{!}{
			\begin{tabular}{m{0.1\textwidth} m{0.84\textwidth} }
				\toprule
				\textbf{Type ID} & \textbf{Sampling Type Description} \\
				\midrule
				Type A & Single sample is enough to derive feature confidently \\
				Type B & Single sample is enough to derive feature somewhat confidently while several samples can improve confidence \\
				Type C & Sampling only if there is a significant change is enough to derive feature confidently \\
				Type D & Sampling at regular intervals during a day are enough to derive feature confidently\\
				Type E & Continuous sampling for the duration of the action is enough to derive feature confidently \\
				Type F & Continuous sampling without limitations is needed to derive feature confidently\\
				\bottomrule
			\end{tabular}
		}
		\caption{Sampling types.}
		\label{table:samplingTypes}
	\end{table}
	
	\subsection{GPS and location}
	The Android API package \textit{android.location} is a comprehensive package that integrates GPS, WiFi APs, and Cell Identity information in a proprietary method to provide an accurate location estimate. The \textit{Location} class in this package exposes methods to provide longitude with \textit{getLongitude()}, latitude with \textit{getLatitude}, the accuracy of the estimate with \textit{getAccuracy()}, the altitude with \textit{getAltitude()}, the bearing with \textit{getBearing()}, and even the speed with \textit{getSpeed()}.
	
	Location data has received a significant amount of attention from the research community in the context of privacy. It is seen as a major threat to individual privacy and at the same time its utility is undeniable as evidenced by the vast number of location-based services available. Krumm\cite{Krumm2009} has outlined some of the threats posed by location data. Someone can infer significant places like home and work, and more recently, Do et al.\cite{Do2014a} were able to reliably characterize 10 categories of places of a person's everyday life, these included home and work as well as friend's home, transportation, friend's work, outdoor sport, indoor sport, restaurant or bar, shopping, holiday. Krumm shows examples of how pseudonymized or anonymized location data can still be used to identify the people in the data. Other information such as mode of transportation (bus, foot, car, etc.), age, work role, work frequency, and even smoking habits can also be inferred from location data. The evidence for the privacy risks of location data is overwhelming. For the GPS sensor, Type A sampling is enough to reveal the location while type C and D is enough for detecting personal and significant places.
	
	\subsection{Telephony}
	The network antenna is used to connect to the cellular network (GSM, edge, HSPA, LTE, etc.). The Android API \textit{android.telephony} package can be used to get information such as the identity of the cell tower which the phone is connected to (Cell ID) and the signal strength to this cell using the method \textit{getAllCellInfo()} from the \textit{TelephonyManager} class. This class can also provide the service state with \textit{getServiceState()}, network type with \textit{getNetworkType}, call state with \textit{getCallState()}, and data state with \textit{getDataState()}.
	
	The Cell ID can be used in conjunction with publicly available data of their locations to localize a person as demonstrated by LaMarca et al.\cite{LaMarca2005}. Although a single sample is usually enough to determine an approximate location, several samples might be needed to increase confidence (Type B sampling). As such, the same privacy threats as location can be applied here. However, even without knowing the location of the Cell IDs, one can infer places such as home and work as done by Yadav et al.\cite{Yadav2014} as well as our own work \cite{Fanourakis2013}. Furthermore, since a person's connection traces to Cell IDs is directly related to the person's location traces, the Cell ID traces can be thought of as a quasi identifier much like location. For this, Type C or D sampling is required.
	
	\subsection{Bluetooth}
	The bluetooth antenna is used to connect to nearby bluetooth devices such as wireless headphones or a smartwatch. The Android API \textit{android.bluetooth.le} package can be used to get a list of nearby bluetooth devices using the \linebreak \textit{startScan(...)} method of the \textit{BluetoothLeScanner} class. This method returns a list of class \textit{ScanResult} which includes the hardware ID of the bluetooth devices with the \textit{getDevice()} method and the signal strength with the \textit{getRssi()} method.
	
	Bluetooth connections to personal devices such as headphones and smartwatch are, in general, unique to each individual, as such, they can be used as identifying information. Bluetooth devices in range (not necessarily connected to) are not as unique but provided that some of those Bluetooth devices are geographically stationary then a frequent Bluetooth device scan can also be used to cruedly localize a person as demonstrated again by LaMarca et al.\cite{LaMarca2005}. Type B sampling is recommended for localization, while Type C or D is required to detect personal or significant places.
	
	\subsection{WiFi Antenna}
	The WiFi antenna is used to connect to WiFi networks. The Android API \textit{android.net.wifi} package can be used to get a list of WiFi access points (APs) using the \textit{startScan()} method from the \textit{WifiManager} class. This method returns a list of class \textit{ScanResult} which include the AP identity in the \textit{BSSID} public field and the signal strength in the \textit{level} public field.
	
	WiFi connections to personal access points (APs) such as someone's home or work, much like Bluetooth, can be unique for each individual. It has also been demonstrated that WiFi APs in range and their signal strength can be used to localize a person by LaMarca et al.\cite{LaMarca2005} and Redzic et al.\cite{Redzic2014} among many others \cite{Liu2007,Vo2016}. The same sampling requirements as Bluetooth apply for WiFi.
	
	\subsection{Touchscreen}
	The touchscreen is the main input method on a smartphone, it is used to select items on the screen, to type text, or other gestures which are out of the scope of this work. The Android API package \textit{android.view} includes the class \textit{View.OnTouchListener} which can be used to capture touch events. For security reasons the location of the touch is only available to the application on the foreground, but the touch event itself can still be useful information.
	
	The dynamics of touch events (time between touches, duration of touch, pressure, etc.) are categorized as \textit{keystroke dynamics} and they have been researched heavily for authentication and user recognition for hardware computer keyboards and more recently for smartphones \cite{Banerjee2012,Unar2014,Jain2016}. Frank et al.\cite{Frank2013} show that touchscreen data like navigational strokes (a subset of keystroke dynamics since they do not include typing) cannot be reliably used for authentication as a standalone but provides useful authentication features nonetheless and using this kind of data for authentication is ultimately feasible. Antal et al.\cite{Antal2015} and Roh et al.\cite{Roh2016} among others\cite{Sitova2016} have shown that keystroke dynamics along with additional features that can be collected on a smartphone (accelerometer, pressure, finger area) can be used to improve the performance of authentication. Continuous sampling for the duration of the keystrokes is required for their detection (Type E sampling).
	
	\subsection{Microphone}
	The microphone is used to capture audio to facilitate a phone call or to record audio. The Android API \textit{android.media} package includes the class \textit{AudioRecord} which can be used to capture the audio from the microphone. For the below mentioned exploitation methods of audio signals, continuous sampling is required for the duration of the action in order to apply the methods (Type E sampling).
	
	The audio of someone speaking can be used to recognize them. Speaker recognition is a well researched topic, low level features like short-term spectrum and mel-frequency cepstral coefficients, voice source feature estimation, formant transitions, prosodic features, and high level features such as lexicon have been used in models like vector quantization (VQ), Gaussian mixture models (GMM), support vector machines (SVM), and neural networks \cite{Kinnunen2010}. More recently, with the advent of deep learning, more complex and robust modeling techniques have emerged \cite{Lei2014,Richardson2015}. Speaker recognition has reached a high enough technological maturity level that it has found commercial applications in automated home assistants such as the Google Home.
	
	Many human activities produce characteristic sounds which can be used to recognize them. Activities such as cooking, brushing teeth, showering, washing hands, urinating, shaving, drinking, etc. have been shown to be recognizable by the sounds they produce by several researchers \cite{Kam2005,Zhan2010,Stork2012}. More impressively, not only can someone recognize the activity of typing on a physical keyboard but also recognize what is being typed solely from the data of a microphone \cite{Asonov2004,Berger2006,Zhuang2009}.
	
	Environmental noise features from audio recordings can be used to identify the location of the recording. Acoustic environment identification (AEI), as it is commonly known, is mostly limited to room or enclosed space environments where the geometry of the room can have noticeable effects on the reverberation of the audio. The main applications of AEI are in audio forensics where an estimation of the reverberation and background noise from a recording can be used to identify the room or even the location inside a room where the audio was recorded \cite{Malik2013,Zhao2013,Patole2016,Delgado-Gutierrez2017}. Prior measurements or estimates of the impulse response of the rooms are required for these methods since they describe how the sound reverberates in that room.
	
	Since the room geometry can affect the audio reverberation patterns of a room, someone could use an audio recording of a sharp noise (like a hand clap) to estimate the impulse response of a room and then estimate the dimensions or even the shape of the room \cite{Tervo2012,Markovic2013,Rajapaksha2016}. These methods are often tested under controlled environments and with specialized audio equipment so it is unclear whether a recording from a smartphone microphone would be sufficient for meaningful results.
	
	\subsection{Camera}
	There are often two cameras on a smartphone, the front facing camera and the main camera on the backside of the phone. They are used to take pictures, video, and to facilitate video calls. The Android API \textit{android.hardware.camera2} package provides the necessary methods to retrieve data from the camera.
	
	Pictures or video from a camera can be used in a several different ways to reveal information about the user even without the use of the file metadata. The most obvious is if the subject of the picture is the user themselves or of people related, in the social sense, to the user. If the subject of the picture is a city, a street, or a landmark, algorithms can be used to match the pictures to a location provided there is a database of prior pictures in that location \cite{Zamir2010,Baatz2010,Weyand2016}. There are also algorithms that can recognize the style of an image and match it to a known photographer \cite{Karayev2013,Thomas2016}. Since a single picture is used in these cases, Type A sampling is enough. Videos can also be used with the aforementioned techniques by treating them as sequences of still images. In addition, analyzing the device movement from a video can also be used to identify the user similar to gait recognition in other behavioural biometric identification schemes \cite{Hoshen2016}. Type E sampling is required for this. 
	
	%	\subsection{CPU Sensors}
	%	cpu stuff
	
	%	\subsection{Battery Sensors}
	%	The \textit{BatteryManager} class in the \textit{android.os} package allows us to access information about the battery such as its level, charging status, temperature, voltage, and current load. The battery level is mainly used to provide feedback to the user while the other information is used for estimating the battery health and provide diagnostics.
	
	\subsection{Environmental and Activity Sensors}
	\label{sensors}
	There is a variety of environment and activity sensors on smartphones. Their data is exposed in the Android API \textit{android.hardware} package with the classes \textit{SensorManager}, \textit{Sensor}, and \textit{SensorEvent}. Each sensor type is assigned an integer identifier constant with an appropriate name. Among these sensors are \textit{software sensors}, that is, sensors that do not have a direct hardware counterpart but are calculated from the outputs of one or more hardware sensors. These sensors do not require any special permissions to be accessed which makes it easy for a rogue application or website to get this data without the user's knowledge.
	
	Many of these sensors are based on microelectromechanical systems (MEMS) technology which has been shown to be vulnerable to sensor fingerprinting \cite{Bojinov2014,Baldini2016,VanGoethem2016,Baldini2017,Baldini2017a}. The accelerometer, gyroscope, magnetometer, and barometer are all based on MEMS technology. The idea behind sensor fingerprinting is that minor manufacturing defects give each sensor a unique output which is composed of the true reading (acceleration, magnetic field strength, etc.) plus the bias caused by the manufacturing defect. This makes it so that someone can discriminate the devices which produce a given sensor output. To achieve this, Type E sampling is required.
	In the sections below we will take a look at each individual sensor for their respective privacy threats which are additional to the aforementioned sensor fingerprinting.
	
	\paragraph{Accelerometer}
	The accelerometer (TYPE\_LINEAR\_ACCELERATION) is a hardware sensor that measures linear acceleration. It's main uses include adjusting the display orientation to match the orientation of the physical display and as a step counter among others. To derive other more interesting information besides the orientation of the device, Type E sampling would be required.
	The accelerometer can be used in a variety of ways to become a threat to one's privacy. It has found uses in indoor localization systems where GPS is not available. Together with gyroscope and/or magnetometer readings it can help to accurately track the movement of a person \cite{Lammel2009,Subbu,Redzic2014}. It is often used for activity recognition as well (sitting, walking, running, biking, cleaning, shopping, sleeping, cooking, etc.) \cite{Brajdic2013,Kwapisz2011,Incel2013,Bayat2014}. Its applications also extend into behavioural biometrics where gait recognition uses the accelerometer to recognize a person based on how they walk or move \cite{Kwapisz2010a,Unar2014,Jain2016}. When coupled with touch event detection it has even been used to detect what is being typed on the touch screen \cite{Miluzzo2012}. Therefore the accelerometer can reveal not only location, but activity patterns throughout one's daily life, the identity of someone based on how they walk, and in some cases, even what they type on their smartphone.
	It has been shown that auditory vibrations can be picked up by the accelerometer on modern smartphones like the iPhone 4 or a Samsung Galaxy S4 and can be used to detect hotwords (short keywords or phrases that are often used to activate voice assistants) or even what is being typed on a physical keyboard nearby \cite{Marquardt2011,Zhang2015a}.
	
	\paragraph{Gyroscope}
	The gyroscope (TYPE\_GYROSCOPE) is a hardware sensor that measures the rotation or twist of the device. It is often used in conjunction with the accelerometer to measure the orientation of the device and to aid in navigation/localization schemes.
	Michalevsky et al.\cite{Michalevsky2014} show that sounds can affect the measurements of a gyroscope to such a level that private information about the phone's environment can be revealed such as who is speaking and to some extent, what is being said. Type E sampling is required for these methods.
	
	\paragraph{Magnetometer}
	The magnetometer (TYPE\_MAGNETIC\_FIELD) is a hardware sensor that is mainly used to measure the Earth's magnetic field for the purpose of navigation.
	It has found uses in indoor localization schemes by comparing the magnetic field to previously collected magnetic field fingerprints to localize a person \cite{Kim2012,LeGrand2012,Subbu2013}. These methods require Type E sampling and prior data collection to map the fingerprint to specific locations. It is not applicable for outdoor environments since these methods rely on the structural supports of building and rooms which produce these magnetic fingerprints. For outdoor environments it can only reliably measure the orientation of the smartphone with respect to the Earth's magnetic field.
	
	\paragraph{Barometer}
	The barometer (TYPE\_PRESSURE) is a hardware sensor that measures the atmospheric pressure. Not all devices are equipped with this sensor.
	Barometric pressure varies depending on the weather and on altitude. Baring extreme weather events, the rate of change of barometric pressure due to weather is relatively slow (less than 0.04hPa per hour for steady weather, less than 0.5hPa per hour for slow weather changes, and up to 3hPa per hour for rapid weather changes). While in a city like Geneva, Switzerland where the highest altitude is 457m and lowest is 370m, one can expect a change of approximately 0.115hPa per meter of altitude change. Based on these crude estimates it is no surprise that the barometric pressure is often used as an altimeter and with its inclusion in smartphones it has aided in indoor navigation algorithms to determine the floor that the person is on \cite{Lammel2009,Xia2015,Ye2016,Falcon2017}. As such, someone with access to barometer data can learn about the altitude or floor in which a person lives and works as well as altitude variations during their commute. The specific methods vary in their sampling from Type B to Type F. For a city with many altitude variations like Geneva, it does not seem out of the realm of possibility to be able to reconstruct the commute path of a person based on barometric data, it is something worth looking into.
	
	\paragraph{Proximity}
	The proximity sensor (TYPE\_PROXIMITY) is a hardware sensor that measures distance. It is mainly used to detect when the user places the device next to their ear during a phone call so that the screen can be turned off in order to save power. In most cases the sensor has a very limited range of up to 5cm and only tells you if there is something near it (less than 5cm). As such, it is only useful to know if the phone is in a pocket, bag, or next to your ear when taking a call. It does not appear to have any immediate implications to privacy.
	
	\paragraph{Ambient light}
	The ambient light sensor (TYPE\_LIGHT) is a hardware sensor that measures the intensity of light. It is mainly used to automatically adjust the screen brightness to a comfortable level. Ambient light during daytime varies significantly for indoor and outdoor locales, therefore, someone can easily detect this during the daytime using this sensor \cite{Zhou2012}. Type C or D sampling would be enough to detect when the user changes from indoor to outdoor throughout the day.
	Kayacik\cite{Kayacik2014} and Micallef et al.\cite{Micallef2015} created temporal and spatial models for light sensor readings among other sensors and their results show that the light sensor readings are among the sensors with the highest similarity between users. Based on their results they conclude that, on its own, the light sensor is not sufficient for authentication. 
	An interesting exploit of the ambient light sensor was revealed by Spreitzer\cite{Spreitzer2014} where they showed that by using variations in the ambient light due to slight tilting of the smartphone while inputting a PIN they can improve their chances of correctly guessing it. They used a corpus of 50 random PINs and allowed themselves 10 guesses and managed to have an 80\% success rate compared to 20\% if they randomly guessed. Type E sampling during the PIN entry was used.
	The ambient light sensor has also found a use in indoor localization. If one has control of the LED lighting in a room they can send detectable light variations to the phone and help it to localize itself in the room \cite{Li2014}. Mazilu et al.\cite{Mazilu2015} have also shown that it is feasible to detect room changes solely based on the ambient light sensor readings. Both of these indoor localization methods require Type E sampling.
	
	\paragraph{Gravity}
	The gravity sensor (TYPE\_GRAVITY) is a software sensor that provides the direction and acceleration due to gravity. It most commonly uses the readings of the accelerometer and the gyroscope. It is directly correlated with the physical orientation of the device. The main use of this software sensor is to remove the gravity component from raw accelerometer measurements and be able to use those measurements for other tasks that require only the linear acceleration. On their own, the gravity measurements have very little utility and therefore do not pose any apparent threat to privacy.
	
	\paragraph{Step}
	The step sensor (TYPE\_STEP\_COUNTER, TYPE\_STEP\_DETECTOR) is a software sensor that detects when the steps a user makes when walking. It uses the accelerometer readings to derive the steps.
	When stride length is known (distance after one step) or accurately estimated from the height of a person, step counts can be used to estimate the distance that a person has walked \cite{Bassett1996,Tudor-Locke2002,Crouter2003,Chon2011}. Since only one sample is needed to derive the distance, Type A sampling is enough. Although there is significant error depending on what device is being used or even depending on the speed that a person is walking, someone can roughly determine the distances to nearby destinations where the user walks to. There are no significant privacy concerns for this data since the accuracy of these measurements can have significant errors over longer distances or even at different walking speeds.
	\clearpage
	
	\subsection{Summary}
	In table \ref{table:sensorThreatSummary} we summarize the possible threats of each sensor noting the type of sampling that is required. Location and location features seem to be a common type of threat for most sensors.
	In table \ref{table:sensorThreatSurvey} we summarize the literature which was used.
	
	{
		\renewcommand{\arraystretch}{1.5}
		\begin{table}[H]
			\resizebox{\textwidth}{!}{
				\footnotesize
				\begin{tabular}{m{0.15\textwidth} m{0.42\textwidth} m{0.3\textwidth}}
					\toprule
					\textbf{Sensor} & \textbf{Threat Summary} & \textbf{Sampling Reqs}\\
					\midrule
					GPS & location and personal places\cite{Krumm2009,Do2014a} & Type A for location, Type C and D for personal places \\
					%\hline
					Cell ID & location and personal places\cite{LaMarca2005,Kang2005,Fanourakis2013} & Type B for location, Type C and D for personal places \\
					%\hline
					Bluetooth & location and personal places\cite{LaMarca2005}, identity (from connections to personal devices)  & Type B for location, Type C and D for personal places and identity\\
					%\hline
					WiFi & location and personal places\cite{LaMarca2005,Liu2007,Redzic2014,Vo2016}, identity (from connections to personal devices) & Type B for location, Type C and D for personal places and identity\\
					%\hline
					Touchscreen & identity (keystroke dynamics\cite{Banerjee2012,Frank2013,Unar2014,Antal2015,Jain2016,Roh2016,Sitova2016}) & Type E \\
					%\hline
					Microphone & identity (speaker recognition\cite{Kinnunen2010,Lei2014,Richardson2015}), activity\cite{Kam2005,Zhan2010,Stork2012}, keylogger (for physical keyboard\cite{Asonov2004,Berger2006,Zhuang2009}), location features (AEI\cite{Malik2013,Zhao2013,Patole2016,Delgado-Gutierrez2017}, room characteristics\cite{Tervo2012,Markovic2013,Rajapaksha2016}) & Type E \\
					%\hline
					Camera & location and location features\cite{Zamir2010,Baatz2010,Weyand2016}, identity (selfies, gait recognition from video\cite{Hoshen2016}, author recognition\cite{Karayev2013,Thomas2016}) & Type A for static pictures, Type E for video \\
					%\hline
					All MEMS & identity (MEMS sensor fingerprinting\cite{Bojinov2014,Baldini2016,VanGoethem2016,Baldini2017,Baldini2017a}) & Type E \\
					%\hline
					Accelerometer (MEMS) & location (indoor navigation\cite{Lammel2009,Subbu,Redzic2014}), activity\cite{Brajdic2013,Kwapisz2011,Incel2013,Bayat2014}, PIN\cite{Miluzzo2012}, identity (gait recognition\cite{Kwapisz2010a,Unar2014,Jain2016}, speaker recognition\cite{Marquardt2011,Zhang2015a}) & Type E \\
					%\hline
					Gyroscope (MEMS)& identity (speaker recognition\cite{Michalevsky2014}) & Type E \\
					%\hline
					Magnetometer (MEMS)& location (indoor localization via fingerprinting\cite{Kim2012,LeGrand2012,Subbu2013}) & Type E\\
					%\hline
					Barometer (MEMS)& location features (floor detection\cite{Lammel2009,Xia2015,Ye2016,Falcon2017}) & Type B up to Type F \\
					%\hline
					Proximity & None  \\
					%\hline
					Ambient light & location features (indoor vs outdoor\cite{Zhou2012}, indoor navigation\cite{Li2014}, room detection\cite{Mazilu2015}), PIN \cite{Spreitzer2014} & Type C and D for indoor/outdoor/room features, Type E for navigation and PIN \\
					%\hline
					Gravity & None  \\
					%\hline
					Step & distance walked (estimated from number of steps \cite{Bassett1996,Tudor-Locke2002,Crouter2003,Chon2011}) & Type A \\
					\bottomrule
				\end{tabular}
			}
			\caption{Summary of sensors and corresponding privacy threats}
			\label{table:sensorThreatSummary}
		\end{table}
	}
	
	\clearpage
	{ 
		\footnotesize
		\begin{longtable}{cm{0.3\textwidth}m{0.4\textwidth}}
			\toprule
			%\rowcolor{lightgray}
			\textbf{Citation} & \textbf{Sensors used} & \textbf{Derived information}\\ 
			%& \textbf{Ground truth} & \textbf{Methodology} & \textbf{Accuracy \& Limitations}\\
			\midrule
			\endhead
			\\
			\multicolumn{3}{c}{{Continued on Next Page\ldots}}
			\endfoot
			\endlastfoot
			\rowcolor{lightgray}
			\cite{Do2014a} & GPS, WiFi, Bluetooth, App & Location of home, work, other personal places\\*
			\multirow{3}{*}{\rotatebox[origin=c]{90}{Details}}
			&\multicolumn{2}{m{0.7\textwidth}}{\textbf{Ground truth:} User annotated data.}\\*
			&\multicolumn{2}{m{0.7\textwidth}}{\textbf{Methodology:} Random forest classifier.}\\*
			&\multicolumn{2}{m{0.7\textwidth}}{\textbf{Accuracy \& Limitations:} GPS features alone gave $70.3\%$ accuracy, adding Wifi features to previous $71.7\%$, adding Bluetooth features to previous $74.6\%$, adding app features to previous $75\%$. Infrequently visited places are not reliably recognized.}\\
			\rowcolor{lightgray}
			\cite{LaMarca2005} & WiFi, Bluetooth, Cell ID & Map of radio beacons, location of user\\*
			\multirow{3}{*}{\rotatebox[origin=c]{90}{Details}}
			&\multicolumn{2}{m{0.7\textwidth}}{\textbf{Ground truth:} GPS war-driving or institution databases with location of radio beacons.}\\*
			&\multicolumn{2}{m{0.7\textwidth}}{\textbf{Methodology:} tracker component that models signal propagation and takes into account physical environment (for example, buildings). A probabilistic Bayesian particle filter can be used to increase accuracy.}\\*
			&\multicolumn{2}{m{0.7\textwidth}}{\textbf{Accuracy \& Limitations:} lower accuracy than GPS.}\\
			\rowcolor{lightgray}
			\cite{Kang2005} & GPS, WiFi, Bluetooth, Cell ID & Location personal places\\*
			\multirow{3}{*}{\rotatebox[origin=c]{90}{Details}}
			&\multicolumn{2}{m{0.7\textwidth}}{\textbf{Ground truth:} user annotated data.}\\*
			&\multicolumn{2}{m{0.7\textwidth}}{\textbf{Methodology:} GPS or PlaceLab estimated location was used to collect traces. Time based clustering was used on location traces to find personal places.}\\*
			&\multicolumn{2}{m{0.7\textwidth}}{\textbf{Accuracy \& Limitations:} Does not label the personal places.}\\
			\rowcolor{lightgray}
			\cite{Fanourakis2013} & Cell ID & Detection of personal place \\*
			\multirow{3}{*}{\rotatebox[origin=c]{90}{Details}}
			&\multicolumn{2}{m{0.7\textwidth}}{\textbf{Ground truth:} GPS and user annotated data.}\\*
			&\multicolumn{2}{m{0.7\textwidth}}{\textbf{Methodology:} graph based clustering of Cell IDs using Cell ID transition matrix populated by Cell ID oscillation events. Duration of stay in clusters and time of day indicating home or work.}\\*
			&\multicolumn{2}{m{0.7\textwidth}}{\textbf{Accuracy \& Limitations:} limited to urban environment with relatively dense cellular tower deployment. Does not detect places with shorter durations of stay.}\\
			\rowcolor{lightgray}
			\cite{Redzic2014} & WiFi & indoor location \\*
			\multirow{3}{*}{\rotatebox[origin=c]{90}{Details}}
			&\multicolumn{2}{m{0.7\textwidth}}{\textbf{Ground truth:} ground truth.}\\*
			&\multicolumn{2}{m{0.7\textwidth}}{\textbf{Methodology:} Fingerprinting of RSSI of WiFi access points at specific calibration points (CPs) and using naive Bayes to identify the three nearest CP, then using interpolation driven by the likelihoods to find the location of the user in the vicinity of those CPs (even using as few as 2 of them).}\\*
			&\multicolumn{2}{m{0.7\textwidth}}{\textbf{Accuracy \& Limitations:} Accuracy is around 2 meters which can be significant in indoor environments even though they showed that this method is better than many others. Requires calibration measurements in advance.}\\
			\rowcolor{lightgray}
			\cite{Frank2013} & touchscreen (navigational strokes) & user identity\\*
			\multirow{3}{*}{\rotatebox[origin=c]{90}{Details}}
			&\multicolumn{2}{m{0.7\textwidth}}{\textbf{Ground truth:} 41 users read text and compare images on an android phone to produce natural navigational strokes.}\\*
			&\multicolumn{2}{m{0.7\textwidth}}{\textbf{Methodology:} 30 behavioural touch features (for example, mid-stroke area covered, direction of end to end line, start/end x, start/end y, and more). From sets of highly correlated features, only one was selected. Used kNN and SVM classifiers.}\\*
			&\multicolumn{2}{m{0.7\textwidth}}{\textbf{Accuracy \& Limitations:}  0\% to 4\% error (false negative and false positive combined) which is not ideal for authentication purposes. More subjects needed to improve feature selection. Differences of screen sizes of devices needs to be taken into account.}\\
			\rowcolor{lightgray}
			\cite{Antal2015} & touchscreen & user identity\\*
			\multirow{3}{*}{\rotatebox[origin=c]{90}{Details}}
			&\multicolumn{2}{m{0.7\textwidth}}{\textbf{Ground truth:} 42 users. Android application with its own keyboard. Nexus 7 tablet (37 users) and LG Optimus L7 II p710 phone (5 users). Users input a password 30 times (same for all).}\\*
			&\multicolumn{2}{m{0.7\textwidth}}{\textbf{Methodology:} features: time between key press and release, time between consecutive key presses, time between key release and next press, pressure of press, finger area of press, averages of previous values. WEKA machine learning software was used. Analyzed several classifiers.}\\*
			&\multicolumn{2}{m{0.7\textwidth}}{\textbf{Accuracy \& Limitations:} Best classifier was random forest with 82.53\% accuracy using only time based features, and 93.04\% accuracy using time based features and touchscreen based features together.}\\
			\rowcolor{lightgray}
			\cite{Sitova2016} & touchscreen, accelerometer, gyroscope, magnetometer & user identity\\*
			\multirow{3}{*}{\rotatebox[origin=c]{90}{Details}}
			&\multicolumn{2}{m{0.7\textwidth}}{\textbf{Ground truth:} 100 users typing 3 answers of at least 250 words under sitting or walking conditions. Sensor sampling at 100Hz.}\\*
			&\multicolumn{2}{m{0.7\textwidth}}{\textbf{Methodology:} Scaled Manhattan (SM), scaled Euclidean (SE), SVM verifiers using hand movement, orientation, grasp (HMOG) features, tap and keystroke dynamics features.}\\*
			&\multicolumn{2}{m{0.7\textwidth}}{\textbf{Accuracy \& Limitations:} Best verifier was SM with Equal Error Rate of 10.05\% for sitting and 7.16\% for walking. Including HMOG features improved accuracy over only tap or keystroke dynamics. Cross-device interoperability and varying walking speeds were not explored.}\\
			\rowcolor{lightgray}
			\cite{Zhan2010} & microphone & Human activity (cleaning, brush teeth, walk, drink water, etc.)\\*
			\multirow{3}{*}{\rotatebox[origin=c]{90}{Details}}
			&\multicolumn{2}{m{0.7\textwidth}}{\textbf{Ground truth:} Sound recordings of each activity}\\*
			&\multicolumn{2}{m{0.7\textwidth}}{\textbf{Methodology:} 5 random segments of 1.5 second from recording were used. Mel Frequency Cepstral Coefficients (MFCC) were extracted for each segment. Discrete time warping was used to get closest match.}\\*
			&\multicolumn{2}{m{0.7\textwidth}}{\textbf{Accuracy \& Limitations:} Average accuracy of recognizing each of the 14 activities was 92.5\% (80\% lowest, 100\% highest). Sound samples were recorded in a controlled environment, realistic data would improve argument.}\\
			\rowcolor{lightgray}
			\cite{Zhuang2009} & microphone & text typed on physical keyboard \\*
			\multirow{3}{*}{\rotatebox[origin=c]{90}{Details}}
			&\multicolumn{2}{m{0.7\textwidth}}{\textbf{Ground truth:} 10 minute recording of user typing in English}\\*
			&\multicolumn{2}{m{0.7\textwidth}}{\textbf{Methodology:} Compute Cepstrum features of each keystroke. For training, use clustering technique to separate into classes and language model correction based on HMM to label and then train a classifier. For recognition, use classifier and language model correction.}\\*
			&\multicolumn{2}{m{0.7\textwidth}}{\textbf{Accuracy \& Limitations:} 90\% of 5-character passwords in fewer than 20 attempts, 80\% of 10-character passwords in fewer than 75 attempts. Classifiers user: linear classification, Gaussian mixtures, or Neural Network.}\\
			\rowcolor{lightgray}
			\cite{Patole2016} & microphone & environment (room)\\*
			\multirow{3}{*}{\rotatebox[origin=c]{90}{Details}}
			&\multicolumn{2}{m{0.7\textwidth}}{\textbf{Ground truth:} 30 audio recordings in 6 different acoustic environments (big classroom 1 and 2, small classroom, small seminar hall, seminar hall, small room)}\\*
			&\multicolumn{2}{m{0.7\textwidth}}{\textbf{Methodology:} Blind de-reverbaration was used to extract reverberant component of audio. Impulse response was estimated via hand-clap method. MFCCs were used as features, a multiclass SVM was used for classification.}\\*
			&\multicolumn{2}{m{0.7\textwidth}}{\textbf{Accuracy \& Limitations:} 4 rooms identified with 100\% accuracy, 2 rooms above 80\% accuracy. Need to measure impulse response of rooms separately. Environments were based only on university campus.}\\
			\rowcolor{lightgray}
			\cite{Markovic2013} & microphone & room dimensions\\*
			\multirow{3}{*}{\rotatebox[origin=c]{90}{Details}}
			&\multicolumn{2}{m{0.7\textwidth}}{\textbf{Ground truth:} simulations of rectangular and L-shaped room.}\\*
			&\multicolumn{2}{m{0.7\textwidth}}{\textbf{Methodology:} Defined a cost function robust against wrong matches of TOAs. Genetic algorithm was used to minimize cost function and derive room dimensions.}\\*
			&\multicolumn{2}{m{0.7\textwidth}}{\textbf{Accuracy \& Limitations:} room dimensions for rectangular room are within 10cm of actual size, for L-shaped room within 70cm. Should be repeated in real room. Room shape known a priori.}\\
			\rowcolor{lightgray}
			\cite{Weyand2016} & camera (photo) & location\\*
			\multirow{3}{*}{\rotatebox[origin=c]{90}{Details}}
			&\multicolumn{2}{m{0.7\textwidth}}{\textbf{Ground truth:} 126M photos with Exif geolocations from the web.}\\*
			&\multicolumn{2}{m{0.7\textwidth}}{\textbf{Methodology:} Used convolutional neural network (CNN) to train with 91M images, the rest used for validation. 237 geotagged Flickr photos used to measure accuracy of model.}\\*
			&\multicolumn{2}{m{0.7\textwidth}}{\textbf{Accuracy \& Limitations:} When using any type of photo accuracy is 8.5\% for 1km radius, 24.5\% for 25km, 37.6\% for 200km, 53.6\% for 750km, 71.3\% for 2500km. Using other contextual info increased accuracy.}\\
			\rowcolor{lightgray}
			\cite{Hoshen2016} & camera (video) & identity\\*
			\multirow{3}{*}{\rotatebox[origin=c]{90}{Details}}
			&\multicolumn{2}{m{0.7\textwidth}}{\textbf{Ground truth:} 32 users recorded two 7 minute sequences with head-mounted cameras.}\\*
			&\multicolumn{2}{m{0.7\textwidth}}{\textbf{Methodology:} optical flow vectors computed for each frame. CNN with 2 hidden layers for classifier.}\\*
			&\multicolumn{2}{m{0.7\textwidth}}{\textbf{Accuracy \& Limitations:} 77\% accuracy for 4 second of video, 90\% accuracy for 12 seconds of video. Stabilizing the video deteriorated results. Requires that camera be mounted on person. Should consider hand-held camera.}\\
			\rowcolor{lightgray}
			\cite{Baldini2016,Baldini2017,Baldini2017a} & MEMS (accelerometer, gyroscope, magnetometer) & device identity\\*
			\multirow{3}{*}{\rotatebox[origin=c]{90}{Details}}
			&\multicolumn{2}{m{0.7\textwidth}}{\textbf{Ground truth:} 3 devices on a robotic arm and moved in a predetermined pattern. For magnetometer, 9 devices were tested, a solenoid was placed around each device and a predetermined signal was produced.}\\*
			&\multicolumn{2}{m{0.7\textwidth}}{\textbf{Methodology:} SVM classifier was used with different kernel functions.}\\*
			&\multicolumn{2}{m{0.7\textwidth}}{\textbf{Accuracy \& Limitations:} Over 95\% accuracy to distinguish between different models, over 65\% accuracy overall. The inputs to the sensors were controlled. This might not be possible to apply with data collected in the wild.}\\
			\rowcolor{lightgray}
			\cite{Kwapisz2011} & accelerometer & human activity (walking, jogging, ascending stairs, descending stairs, sitting, standing)\\*
			\multirow{3}{*}{\rotatebox[origin=c]{90}{Details}}
			&\multicolumn{2}{m{0.7\textwidth}}{\textbf{Ground truth:} 29 users performing each activity several times while carrying a smartphone.}\\*
			&\multicolumn{2}{m{0.7\textwidth}}{\textbf{Methodology:} Split data into 10 second segments, each segment extracted features like average acceleration, standard deviation, time between peaks, etc. WEKA with decision trees (J48), logistic regression, multilayer neural networks (NN) with default settings.}\\*
			&\multicolumn{2}{m{0.7\textwidth}}{\textbf{Accuracy \& Limitations:} NN is best with an average of 91.7\% accuracy. Up/down stairs had the worst accuracy as low as 44.3\% and were most often confused with each other or walking. Activity set is limited, different carrying patterns of device not taken into account.}\\
			\rowcolor{lightgray}
			\cite{Miluzzo2012} & accelerometer, gyroscope & smartphone keyboard input\\*
			\multirow{3}{*}{\rotatebox[origin=c]{90}{Details}}
			&\multicolumn{2}{m{0.7\textwidth}}{\textbf{Ground truth:} 10 users using a custom application for tapping icons and typing text (each letter 50 times, 19 different pangrams, and 20 times the same pangram).}\\*
			&\multicolumn{2}{m{0.7\textwidth}}{\textbf{Methodology:} Detect taps and extracts features of each tap (time domain and frequency domain). kNN, multinomial logistic regression, SVM, random forests, bagged decision trees are all used together in an ensemble classifier.}\\*
			&\multicolumn{2}{m{0.7\textwidth}}{\textbf{Accuracy \& Limitations:} 90\% accuracy for inferring tap locations, 80\% accuracy for letters. The classifier is resource heavy and could have redundancies.}\\
			\rowcolor{lightgray}
			\cite{Marquardt2011} & accelerometer & text typed on physical keyboard\\*
			\multirow{3}{*}{\rotatebox[origin=c]{90}{Details}}
			&\multicolumn{2}{m{0.7\textwidth}}{\textbf{Ground truth:} iPhone placed on same surface as keyboard. Sentences typed were selected from the Harvard Sentences corpus.}\\*
			&\multicolumn{2}{m{0.7\textwidth}}{\textbf{Methodology:} Features from keypress data were used like root mean square, skewness, variance, kurtosis, FFT, MFCCs. Two neural networks were trained with with a difference in features used.}\\*
			&\multicolumn{2}{m{0.7\textwidth}}{\textbf{Accuracy \& Limitations:} Tested with and without dictionary knowledge and with a news article from a newspaper. As much as 80\% accuracy of typed content with the use of dictionary. Orientation of device, desk surface material, typing speed, ambient vibrations can affect the performance.}\\
			\rowcolor{lightgray}
			\cite{Zhang2015a} & accelerometer & hotword detection (for example, "okay Google")\\*
			\multirow{3}{*}{\rotatebox[origin=c]{90}{Details}}
			&\multicolumn{2}{m{0.7\textwidth}}{\textbf{Ground truth:} 10 users recorded saying "Okay google" and 20 common short phrases 10 times each. Each recording played through phone speakers 10 times for training at 70dB.}\\*
			&\multicolumn{2}{m{0.7\textwidth}}{\textbf{Methodology:} 2 second window is used and time domain and frequency domain features are extracted. For mobile scenario, a high pass filter with 2Hz cutoff is used to remove effects due to movement. A decision tree classifier is used.}\\*
			&\multicolumn{2}{m{0.7\textwidth}}{\textbf{Accuracy \& Limitations:} 85\% in static scenario, 80\% in mobile scenario. The mobile scenario is very limited with just a controlled walking. More complicated movements make it vastly more difficult to recognize the hotwords.}\\
			\rowcolor{lightgray}
			\cite{Michalevsky2014} & gyroscope & identity, speech \\*
			\multirow{3}{*}{\rotatebox[origin=c]{90}{Details}}
			&\multicolumn{2}{m{0.7\textwidth}}{\textbf{Ground truth:} Nexus 4, Nexus 7, Samsung Galaxy S III were used. A loudspeaker at 75dB. TIDIGITS corpus was used (recordings of 10 users speaking the 11 digits twice each).}\\*
			&\multicolumn{2}{m{0.7\textwidth}}{\textbf{Methodology:} 10-30ms sliding windows with time domain features and MFCCs and Short Time Fourier Transform (STFT). SVM, GMM, DTW were used as classifiers.}\\*
			&\multicolumn{2}{m{0.7\textwidth}}{\textbf{Accuracy \& Limitations:}  Over 80\% accuracy for gender ID using SVM. Speaker ID \textasciitilde50\% accuracy using DTW with Nexus 4 but 17\% with Samsung. Speaker-independent word rec performed poorly, but improved to 65\% using DTW with speaker-specific models. Results varied significantly between devices.}\\
			\rowcolor{lightgray}
			\cite{LeGrand2012} & magnetometer & location (indoor) \\*
			\multirow{3}{*}{\rotatebox[origin=c]{90}{Details}}
			&\multicolumn{2}{m{0.7\textwidth}}{\textbf{Ground truth:} Magnetic field map data collected by following serpentine pattern in a room on x-axis and then on y-axis. Test data collected following a well defined straight line, or circle path.}\\*
			&\multicolumn{2}{m{0.7\textwidth}}{\textbf{Methodology:} Magnetic field map data was used to generate a map of the field in the room. Test data was then matched to the map using a particle filter.}\\*
			&\multicolumn{2}{m{0.7\textwidth}}{\textbf{Accuracy \& Limitations:} Within 0.7m of ground truth. Wi-Fi was used to get coarse location as initial condition for particle filter.}\\
			\rowcolor{lightgray}
			\cite{Subbu2013} & magnetometer & location (indoor) \\*
			\multirow{3}{*}{\rotatebox[origin=c]{90}{Details}}
			&\multicolumn{2}{m{0.7\textwidth}}{\textbf{Ground truth:} 2 users with HTC Nexus One phone. Magnetic fingerprints collected in hallways of campus buildings as users walked along the walls and pillars.}\\*
			&\multicolumn{2}{m{0.7\textwidth}}{\textbf{Methodology:} Magnetic field fingerprints were collected and then DTW was used on test data to match to the fingerprints.}\\*
			&\multicolumn{2}{m{0.7\textwidth}}{\textbf{Accuracy \& Limitations:} Hallways were detected with over 90\% accuracy after only less than 5 meters of walking. Users were instructed to walk close to objects that influence magnetic fields like pillars. }\\
			\rowcolor{lightgray}
			\cite{Falcon2017} & barometer & location (floor in building)\\*
			\multirow{3}{*}{\rotatebox[origin=c]{90}{Details}}
			&\multicolumn{2}{m{0.7\textwidth}}{\textbf{Ground truth:} 63 trials at 5 different tall buildings in New York City where barometric pressure was recorded and random floors were selected. The user could choose either the staircase or elevator.}\\*
			&\multicolumn{2}{m{0.7\textwidth}}{\textbf{Methodology:} Calculated the change in height based on the international pressure equation. To resolve to a floor number they used calculated clusters derived from data of all visits to building (floor height could be estimated).}\\*
			&\multicolumn{2}{m{0.7\textwidth}}{\textbf{Accuracy \& Limitations:} 65\% accuracy when floor height is not known and a default 4.02m was used (98\% within 1 floor), 100\% accuracy if floor height has been previously estimated.}\\
			\rowcolor{lightgray}
			\cite{Mazilu2015} & ambient light sensor & location (indoor room detection) \\*
			\multirow{3}{*}{\rotatebox[origin=c]{90}{Details}}
			&\multicolumn{2}{m{0.7\textwidth}}{\textbf{Ground truth:} 3 users with Samsung Galaxy S4 collected data in their homes. Users logged room label each time they entered a new room on paper-based diary. Total of 132 hours of data}\\*
			&\multicolumn{2}{m{0.7\textwidth}}{\textbf{Methodology:} If sensor data feature was higher than a fixed threshold then a room change was detected. Decision trees (C4.5) were used for room identification.}\\*
			&\multicolumn{2}{m{0.7\textwidth}}{\textbf{Accuracy \& Limitations:} Using only light sensor, accuracy was around 50\%, with additional sensors like temperature and humidity the accuracy was above 60\%. Random guess was 25\% accuracy at best. Time of day, weather, and open windows affected the performance.}\\
			\rowcolor{lightgray}
			\cite{Spreitzer2014} & ambient light sensor & PIN \\*
			\multirow{3}{*}{\rotatebox[origin=c]{90}{Details}}
			&\multicolumn{2}{m{0.7\textwidth}}{\textbf{Ground truth:} Samsung Galaxy SIII was used. 29 test runs by 10 users who entered 15, 30, or all 50 of the random PINs from 3 to 10 times.}\\*
			&\multicolumn{2}{m{0.7\textwidth}}{\textbf{Methodology:} Multiclass logistic regression, discriminant analysis, and K-nearest neighbor methods were used on the collected data with only light intensity and with additional RGBW information which modern light sensors include.}\\*
			&\multicolumn{2}{m{0.7\textwidth}}{\textbf{Accuracy \& Limitations:} 80\% success after 10 guesses from a set of 50 PINs. The set of 50 PINs is unrealistic as there many more possible combinations. }\\
			\\
			\bottomrule
			\caption{Summary of selected state of the art}
			\label{table:sensorThreatSurvey}
		\end{longtable}
	}
	
	\section{Discussion}
	After reviewing each data type in section \ref{sec:dataThreats} we conclude that most of them can be used on their own to reveal something about a user be it small (for example, the floor on a building) or big (for example, the location of their home and work). Combining different data types can enhance the precision, or accuracy, or both as evidenced by several of the surveyed research in table \ref{table:sensorThreatSurvey}.
	
	On the Android OS there is a permission framework to enable an application to explicitly request from the user if a certain data type can be used or not. Permissions that have a protection level of \textit{normal} are automatically granted by the system while those that have protection level \textit{dangerous} require the user's explicit permission to be allowed. In the older versions of the Android OS, the permissions were requested in batch when installing an application but on the latest Android OS version the permissions are requested individually and on an as-needed basis during the runtime of the application (i.e. until the application needs to use the microphone it will not ask for permission). Furthermore, on the latest Android OS version, a user can adjust the individual data type permissions in the settings for each application after the fact. Consequently, the user is informed about the various data types that an application uses. The permission framework does not cover the sensors in section \ref{sensors} at this moment and it is unclear if it will in the future.
	
	\begin{table}[h]
		\resizebox{\textwidth}{!}{
			\begin{tabular}{lllm{0.17\textwidth}}
				\toprule
				\textbf{Data type} & \textbf{permission} & \textbf{prot. level} & \textbf{comments}\\
				\midrule
				Location (precise) & ACCESS\_FINE\_LOCATION & dangerous & \\
				%\hline
				Location (approximate) & ACCESS\_COARSE\_LOCATION & dangerous & \\
				%\hline
				Network Cell ID & ACCESS\_COARSE\_LOCATION & dangerous & \\
				%\hline
				Bluetooth APs & BLUETOOTH\_ADMIN & normal & \\
				%\hline
				WiFi APs & ACCESS\_WIFI\_STATE & normal & \\
				%\hline
				Touchscreen & No permissions are required & N/A & Touch event only outside application window or touch location available only to the application in foreground \\
				%\hline
				Microphone & RECORD\_AUDIO & dangerous & \\
				%\hline
				Camera & CAMERA & dangerous & \\
				%\hline
				Accelerometer & No permissions are required & N/A & \\
				%\hline
				Gyroscope & No permissions are required & N/A &  \\
				%\hline
				Magnetometer & No permissions are required & N/A &  \\
				%\hline
				Barometer & No permissions are required & N/A &  \\
				%\hline
				Proximity & No permissions are required & N/A & \\
				%\hline
				Gravity & No permissions are required & N/A &  \\
				%\hline
				Step & No permissions are required & N/A &  \\
				\bottomrule
			\end{tabular}
		}
		\caption{Data types and corresponding Android OS permission requirements and protection level}
	\end{table}
	
	Christin et al.\cite{Christin2011a} summarize countermeasures to several privacy threats: tailored sensing and user preferences, pseudonymity, spatial cloaking, hiding sensitive locations, data perturbation, data aggregation, among others. They also present important research challenges in this field that have yet to be fully addressed.
	
	This and most such privacy research are concerned with threats in the context of data collection campaigns for research and data mining but similar principles must also be applied to commercially available smartphone applications. Each year Google has to remove more and more malicious applications from their marketplace amounting to hundreds of thousands of applications\cite{Welch2018,Sulleyman2018}. Although a lot of malicious applications are automatically filtered, some may still slip through and become available for millions of people to install. Some of these can easily collect data such as location from unsuspecting users even if they have to request the specific permission from the user. So many applications require the location permission that a user might not think twice about allowing it.
	In table \ref{table:topApps} we list $25$ of the top installed Android applications\cite{androidrank} along with some of the relevant permissions that are required to fully operate them\cite{googleplay}; the location permission is very common.
	\begin{table}[h]
		\resizebox{\textwidth}{!}{
			\begin{tabular}{lll}
				\toprule
				\textbf{Application} & \textbf{Category} & \textbf{Permissions} \\
				\midrule
				Facebook & Social & Location, Camera, Microphone, WiFi \\
				%\hline
				WhatsApp & Communication & Location, Camera, Microphone, WiFi \\
				%\hline
				Messenger (Facebook) & Communications & Location, Camera, Microphone, WiFi \\
				%\hline
				Subway Surfers & Game Arcade & WiFi \\
				%\hline
				Skype & Communication & Location, Camera, Microphone, WiFi \\
				%\hline
				Clean Master & Tools & Location, Camera, Microphone, WiFi \\
				%\hline
				Security Master & Tools & Location, Camera, Microphone, WiFi \\
				%\hline
				Candy Crash Saga & Game Casual & WiFi \\
				%\hline
				UC Browser & Communication & Location, Camera, Microphone, WiFi \\
				%\hline
				Snapchat & Social & Location, Camera, Microphone, WiFi \\
				%\hline
				My Talking Tom & Game Casual & Microphone, WiFi \\
				%\hline
				Twitter & News \& Magazines & Location, Camera, Microphone, WiFi \\
				%\hline
				Viber Messenger & Communication & Location, Camera, Microphone, WiFi \\
				%\hline
				LINE & Communication & Location, Camera, Microphone, WiFi \\
				%\hline
				Pou & Game Casual & Microphone, WiFi \\
				%\hline
				Super-Bright LED Flashlight & Productivity & Camera \\
				%\hline
				Temple Run 2 & Game Action & WiFi \\
				%\hline
				SHAREit & Tools & Location, Camera, WiFi \\
				%\hline
				imo free video calls and chat & Communication & Location, Camera, Microphone, WiFi \\
				%\hline
				Microsoft Word & Productivity & Camera, WiFi \\
				%\hline
				Flipboard: News For Our Time & News \& Magazines &  \\
				%\hline
				Clash of Clans & Game Strategy & WiFi \\
				%\hline
				Spotify Music & Music \& Audio & Camera, Microphone, WiFi \\
				%hline
				Shadow Fight 2 & Game Action & WiFi \\
				%\hline
				Pokemon GO & Game Adventure & Location, Camera \\
				\bottomrule
			\end{tabular}
		}
		\caption{\textnormal{Top downloaded apps excluding pre-installed and system applications. Only the following permissions were noted: Location, Camera, Microphone, and WiFi.}}
		\label{table:topApps}
	\end{table}
	
	Users should always question if an application really needs a specific permission to function. For example, location can be used for navigation, to check-in to places in social media, to show local weather, to share your location with a contact, for fitness tracking, to show location-based notifications, and many more. The issue arises when an application does not need a precise location (for example, a weather application) or only needs some tracking information (for example, a fitness application). A user should not need to give more information than is needed for an application to function much like when a stranger asks for your contact information you have the choice of giving them any of the following information depending on the intimacy level: first name, last name, email address, phone number, home address, work address, friend's address, parents' address, frequented bars, frequented shops, parents' first and last names, etc. Technically, all of these pieces of information can be considered \textit{contact information} but you would not give out all of them if you only need to give out a first name and an email address for example. Sharing more than is necessary can feel highly intrusive. Therefore, a weather application should only get a meteorological region instead of exact coordinates, and a fitness application should only get distance and speed instead of the exact coordinates of the path you ran. There is currently no mechanism on Android OS or any other popular smartphone operating system that provides this level of abstraction when it comes to location information, location context information, or most other types of data that can be collected on a smartphone device.
	
	\section{Implications for GDPR Compliance}
	The GDPR applies to any entity that handles, uses, or collects "personal data". In the GDPR, personal data is loosely defined as "\textit{any information relating to an identified or identifiable natural person ('data subject'); an identifiable natural person is one who can be identified, directly or indirectly, in particular by reference to an identifier such as a name, an identification number, location data, an online identifier or to one or more factors specific to the physical, physiological, genetic, mental, economic, cultural or social identity of that natural person}". This definition is general enough to cover a wide range of data but for non-experts it is not immediately obvious which data is \textbf{actually} covered.
	
	Each EU member state is responsible for enforcing the GDPR rules by appointing a supervisory authority (SA) who  works with other member state SAs to keep consistency among them. The European Data Protection Board coordinates the SAs. Individuals can submit GDPR claims to the relevant SA who will evaluate the claim and proceed with the appropriate actions. The SA also provides some basic guidance to businesses and organizations in order to help them comply with the GDPR, for example, in the form of a self-assessment checklist. Still, no specific definition is given for what constitutes personal data.
	
	Regarding this issue, in this work, we sought to bring some clarity for smartphone sensor data, a rather small subset of all the possible data types out there. We saw that most of the sensor data on a smartphone can reveal personal information but we can get a sense of how sensitive they are by using the sampling types that we defined (see table \ref{table:samplingTypes}) and assigned to each data type. Data that require a sampling type \textit{A} being very privacy-sensitive and data that require a sampling type \textit{F} being less privacy sensitive. The main implication of the sampling type categorization is that if an organization collects a certain data type at a lesser sampling type than is required to derive sensitive information (see table \ref{table:sensorThreatSummary}), then they may not have to consider this data as personal.
	
	The GDPR has provisions for certification and certification bodies (articles 42 and 43) and a list of these can be found on the European Data Protection Board website\cite{edpb}. At the time of writing of this paper there is no official GDPR certification mechanism but we can expect that there will be in the future. A GDPR certification (along with their seals and marks) can let individuals know about the GDPR compliance of an organization and reassure them that their data is handled accordingly and that they are afforded a certain control over their data. However, such a certification would not inform an individual about the level of potential privacy loss they might incur in the case of a data breach against the organization for example. 
	
	We can use the results this paper to aid in evaluating the potential privacy loss of a set of smartphone sensor data. One way to go about this is to assign numerical values to the threat posed by some type of personal information and then combine it with the sampling type required to derive this personal information. For example, lets suppose that a smartphone application only collects Cell ID data. Location and personal places can be derived from Cell ID data, so we will assume that it is a high privacy threat. The sampling type required to derive this information is \textit{C}. Combining the two measures should result in a relatively high potential privacy loss. On the other hand, lets consider barometer data. This data, sampled at the sampling type \textit{B} level, can be used to reveal the altitude or floor level of an individual which seems like a low privacy threat. This should result in a low overall potential privacy loss. The specifics of this evaluation can be further investigated in the future and a concrete methodology can be established for all data types, not just smartphone sensor data, and organizations can display this score in an effort to inform the users about the level of privacy loss that a user might incur in a worst case scenario.
	
	\section{Future Work}	
	In order to have a complete summary of privacy threats from all mobile device data types, more data sources need to be investigated. In this paper we only looked at the most commonly used sensor data, but there are other sensors such as the CPU temperature and battery state which also warrant investigation. Furthermore, there are many software sources that need to be scrutinized. Some of these include application usage, phone interaction (for example, screen on/off), browsing history, instant messaging behaviour, TCP connection information, and more.

	\bibliography{PaperReferences-pii_paper1}
		
\end{document}